\documentclass[pra,twocolumn,a4paper,superscriptaddress,amsmath,amssymb,floatfix]{revtex4-1}
\usepackage{graphicx,epstopdf,xcolor}
\usepackage{enumerate}
\usepackage{braket}
\usepackage[colorlinks]{hyperref}
\hypersetup{%
  linkcolor=blue,
  citecolor=blue,
  filecolor=black,
  menucolor=black,
  urlcolor =black
}

\newcommand{\change}[1]{\textcolor{black}{#1}}

\def\openone{\leavevmode\hbox{\small1\kern-3.3pt\normalsize1}}

\begin{document}

\author{J. Martin Berglund}
\affiliation{Theoretische Physik, 
Universit\"{a}t  Kassel, Heinrich-Plett-Stra{\ss}e 40,
34132 Kassel, Germany}
\affiliation{Instituto de física, 
Universidad Autónoma de San Luis Potosí, UASLP, Av. Parque Chapultepec 1570, Privadas del Pedregal, 78295 San Luis Potosí, S.L.P.}

\author{Michael Drewsen}
\affiliation{Department of Physics and Astronomy, 
Aarhus University, Ny Munkegade 120, DK-8000 Aarhus, Denmark}

\author{Christiane P. Koch}
\email{E-mail: christiane.koch@fu-berlin.de}
\affiliation{Theoretische Physik, 
Universit\"{a}t  Kassel, Heinrich-Plett-Stra{\ss}e 40,
34132 Kassel, Germany}
\affiliation{Freie Universit\"{a}t Berlin, 
Fachbereich Physik \& Dahlem Center for Complex Quantum Systems, Arnimalle 14, 14195 Berlin, Germany}

\title{Rotational excitation in sympathetic cooling of diatomic molecular ions
by laser-cooled atomic ions}

\date{\today}

\begin{abstract}
Sympathetic cooling of molecular ions through the Coulomb interaction
with laser-cooled atomic ions is an efficient tool to prepare
translationally cold molecules 
\change{without, ideally, affecting the internal state of the molecular ions.
However, the electric field due to the Coulomb interaction may}
induce rotational transitions that change the purity of initially
quantum state prepared molecules. Here, we use estimates of rotational
state changes in single collisions of diatomic ions with atomic ions [arXiv:1905.02130]
to determine the overall rotational excitation accumulated over the
sympathetic cooling. \change{Considering two different experimental scenarios, that of a molecular ion co-trapped with} a single atomic ion and \change{a molecular ion immersed in} a Coulomb crystal of atomic ions, we also estimate the cooling time.
\end{abstract}

\maketitle

\section{Introduction}
Sympathetic cooling of molecular ions by means of co-trapped laser-cooled atomic ions~\cite{MolhavePRA00, Drewsen99} is an efficient method for creating cold and ultra-cold molecular ions. Since cooling is mediated by the mutual Coulomb repulsion between the ions, it is independent of the details of the internal structure of the ions and therefore generally applicable. Consequently, it has been applied to a wide range of different species, including various  polar~\cite{MolhavePRA00,KoelemeijPRL07,StaanumPRL08,WillitschPRL08,HansenAngewandte12}
and apolar~\cite{BlythePRL05,TongPRL10} diatomic molecular ions as well as polyatomic molecular ions~\cite{OstendorfPRL06,HojbjerrePRA08,CalvinPRA23}, some of which have been successfully cooled down to temperatures of some tens of milli-Kelvin. The study of molecular ions is currently \change{an active research topic} with applications in e.g., quantum logic protocols~\cite{Drewsen13,Sin22b,FureyQuantum24}, fundamental physics~\cite{schiller05,SafronovaRMP18,HutzlerQST20,ErezPRX23} and cold and ultra-cold chemistry~\cite{OsterwalderBook2018,Deiss24,CalvinPRL25}.

In particular, \change{rovibrationally} cold molecular ions are a versatile tool on which to implement quantum technologies~\cite{Deiss24}. Polar molecular ions benefit from their permanent dipole interaction, which allows for effective control over the molecular ion and has been used to implement various quantum control protocols~\cite{schuster2011cavity, Drewsen13, lin2020quantum, chou2017preparation}. A drawback with polar molecular ions is that they may couple to the black body radiation via their dipole moments. A way around this issue is to utilize apolar molecular ions~\cite{Willitsch20}, which lack a permanent dipole moment. 
\change{The pristine control over the molecular ions has allowed to leverage quantum logic spectroscopy~\cite{Piet05} for molecular quantum state control with unprecedented spectral resolution~\cite{WolfNature16,Willitsch20}. }
Molecular ions can also be combined into hybrid systems with neutral species for a wide range of applications in quantum technology and fundamental physics~\cite{Deiss24}. 

Cooling of both the translational and internal degrees of freedom is of great interest in this respect. Indeed, sympathetic cooling by one laser-cooled atomic ion can cool one molecular ion to the ground state of the ion pair within the trap~\cite{WanPRA15,RugangoNJP15,WolfNature16,ChouNature17,PoulsenPhD}. 
Simultaneously, optical pumping~\cite{VogeliusPRL02,StaanumNatPhys10,SchneiderNatPhys10,DebPCCP13}, helium buffer gas cooling~\cite{HansenNature14}, probabilistic state preparation~\cite{WolfNature16,ChouNature17} and resonance enhanced multi-photon ionization (REMPI)~\cite{TongPRL10,GardnerSciRep19} have been applied to produce molecular ions in specific internal states with high probability.

The latter method differs from the others in that the internal quantum state is prepared \textit{prior} to sympathetic translational cooling and may thus be prone to state changes during cooling. So far, state-selective REMPI has only been demonstrated to form state-selected N$_2^+$ molecular ions in the vicinity of the trap potential minimum~\cite{TongPRL10}. \change{For other molecules, however, this may not be} practical, for example, when the neutral precursor molecule (such as H$_2$ or HD) reacts efficiently with the laser-cooled atomic ions. In such cases, it can be favorable to first produce the state-selected molecular ions in one trap and transfer them into another trap more suitable for translational sympathetic cooling. A similar situation is encountered for molecular ions that have to be produced by an external source, such as an electrospray ion source combined with an internal state pre-cooling trap~\cite{StockettRSI16}. In all these cases, the initial kinetic energy of the captured molecular ions can 
be as high as the effective trap potential, typically in the $1-10\,$eV range.

A potential issue with sympathetic cooling of molecular ions arises from the
Coulomb field, originating from the atomic coolants. \change{It} may also couple to the rotational degree of freedom and thereby induce rotational state changes during the scattering events of the cooling process~\cite{BerglundPRA1}. 
\change{The excitation probability depends sensitively on the molecular parameters as well as the collision energy~\cite{BerglundPRA1}. Sympathetic cooling typically requires multiple collisions to extract the translational energy, and each of these collisions occurs at a different collision energy. Changes in the internal molecular state, each accompanied by a loss of purity, may thus accumulate. This raises the question whether sympathetic cooling induces a significant error in the internal molecular state at the end of the cooling process. This is the question we address here.}
\change{To this end,} we consider sympathetic cooling of diatomic molecular ions with \change{an emphasis} on the translational cooling dynamics and rotational state excitations. We \change{also} investigate the feasibility of cooling depending of the interatomic distances of the atomic ions in the trap \change{which effectively set the rate of collisions.} 

\change{The starting point of our investigation is the rotational excitation probability for a given collision energy that has been determined by leveraging the separation in time, resp. energy scales, of translational (relative) and rotational (internal) motion~\cite{BerglundPRA1}. Using a classical description of the translational motion and approximating both atomic ion and molecular ion as point particles, the inter-particle separation is given by Kepler's law of motion, as known from textbook classical mechanics. The electric field due to the Coulomb interaction also couples to the internal molecular degrees of freedom. While vibrational excitations can be neglected, the electric field may induce rotational state changes. These have been calculated by solving the time-dependent Schrödinger equation of a molecular rotor in a time-dependent electric field, where the time-dependence arises from the relative motion during the collision. Incidentally, the time-dependent field has a nearly Lorentzian profile, allowing for analytical approximations of the rotational dynamics~\cite{BerglundPRA1}.}
\change{Due to their fundamentally different couplings to an electric field, polar and apolar molecular ions have been treated separately. Polar species possess a permanent dipole moment, and the interaction is dominated by the dipole term. Apolar species on the other hand, lack a permanent dipole moment by symmetry, and the leading order contributions are due to the induced dipole moment (or polarizability anisotropy) and the quadrupole moment. Numerical simulations of the rotational dynamics revealed two different regimes for polar and apolar molecular ions~\cite{BerglundPRA1}: For apolar species, the quadrupole interaction is found to dominate, and population excitation remains small throughout the collision, allowing for a reasonably accurate analytical approximation at the end of the collision via first order perturbation theory. This closed-form estimate depends on the collision energy and the impact parameter, in addition to the atomic mass and molecular parameters. 
In contrast, for polar molecular ions, the dipole interaction, which is linear in the electric field strength and varies only gradually at large distances, results in a slow alignment (de-alignment) as the two particles approach (veer away from) each other~\cite{BerglundPRA1}. This in turn allows for the dynamics of a single collision to be analyzed in the adiabatic picture. No closed expression for the excitation probability was found for collisions proceeding in the high field limit with strong following of the field. But for molecular ions with a small dipole moment, the effect of the field on the rotational states is moderate, and an approximate, two-level model can be used. This yields an upper bound on the excitation probability, that again depends on the collision energy and impact parameter. In particular, all relevant rotational excitation occurs for impact parameters on the order of several hundreds of Bohr or less~\cite{BerglundPRA1}.}

\change{With this comprehensive understanding of rotational state changes in a single collision of a molecular ion with an atomic ion, we are now in the position to estimate the accumulated rotational state change at the end of a sympathetic cooling process involving many such collisions. To determine how many collisions are necessary to reach, from a given initial energy, a desired final kinetic energy of the molecule, we need to estimate the energy transfer per collision. Knowing the energy transfer, we can then construct the sum over all cooling steps to calculate the overall probability for rotational state changes as well as the overall cooling time.}

\begin{figure}[tb]
 \centering
\includegraphics[width=0.79\linewidth]{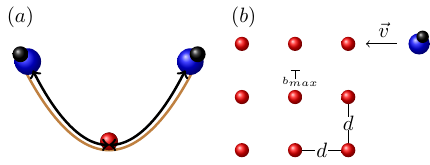}
 \caption{Sympathetic cooling of a molecular ion via collisions with  laser-cooled atomic ions: 
The considered cooling scenarios with either a single trapped atom (a) or many atomic ions forming a Coulomb crystal (b).}
 \label{fig:scheme}
\end{figure}

\change{Just as the population excitation, the energy transfer in a single collision depends on the collision energy and impact parameter. 
The latter is the perpendicular distance between the initial velocity vector of the molecular ion and the center of the Coulomb potential, i.e., the atomic ion; it quantifies how "off-center" the collision is before the particles begin to interact~\cite{Goldstein}. While the collision energy is determined by the energy transfer in the previous collision, the impact parameter is not controllable. We account for this by averaging all relevant quantities over the impact parameter. 
The idea is that the molecular ion is injected into an ion trap holding a single laser-cooled atomic ion or a laser-cooled Coulomb crystal of many atomic ions, as depicted in Fig.~\ref{fig:scheme}. The maximum distance of the molecular ion from an atomic ion then determines the range of possible values that the impact parameter can take, as indicated in the figure. 
The problem of finding the energy transfer and construct the averages over the impact parameter for the two scenarios of Fig.~\ref{fig:scheme} are the subject of the present study, whereas the excitation probabilities at a given scattering energy and impact parameter are reported in the companion paper~\cite{BerglundPRA1}. }

\change{The remainder of this paper is organized as follows. We describe the averaging over all relevant impact parameters and derive the amount of kinetic energy transferred in a single collision in Sec.~\ref{sec:translational-cooling}. This allows us to estimate the overall cooling time for the two scenarios of Fig.~\ref{fig:scheme}. We then present the results for the rotational excitation at the end of the sympathetic cooling process in Sec.~\ref{sec:rot-excitation} for both apolar and polar molecules. We conclude with an outlook in Sec.~\ref{sec:concl}.
}

\section{Translational cooling}
\label{sec:translational-cooling}

Our model is built on a separation of energy scales for translational and internal
molecular motion: While initial collisional energies range from $0.1\,$eV to 10$\,$eV,  the rotational energy scale is only of the order of $10^{-4}\,$eV, so we can treat the translational dynamics without having to worry about changes in collisional energies due to rotational excitations and vice versa~\cite{BerglundPRA1}. 
We inspect two sympathetic cooling regimes, illustrated in Fig.~\ref{fig:scheme}(a) and~(b), respectively. In both cases, we assume the ion trap potential to be harmonic and isotropic, and either contain a single atomic ion~\cite{StaanumPRL08,HansenAngewandte12,TongPRL10} (a) or many atomic ions in the form of a Coulomb crystal~\cite{StaanumNatPhys10,HeazlewoodARPC15} (b), which for simplicity is assumed to have a simple cubic structure (though often fcc or bcc). As we shall see \change{at the end of this section}, the specific scenario, \change{(a) or (b),} will be important for the time required to reach the final energy.

Treating the molecular ion as a point particle \change{as far as the translational motion is concerned}, the relative \change{translational} motion reduces to the textbook problem of classical scattering in a
$1/r$-potential~\cite{Goldstein}. This neglects the trap potential which is reasonable at the relevant short distances where effective translation energy exchange take place~\cite{BerglundPRA1}. 
Classical scattering is characterized by the collision energy $E$ and the impact parameter $b$ as well as the reduced mass $\mu = m_{mol}M_{at}/(M_{mol}+M_{at})$  with $M_{mol}$ and $M_{at}$ the molecular and atomic masses.
\change{Since t}he impact parameter is not fixed, nor controllable, in a sympathetic cooling experiment\change{, we } need to average over all possible values of $b$.
In the following, we focus on two extreme situations---cooling by a single atom (SA) and a large Coulomb crystal (CC), cf. Fig.~\ref{fig:scheme}.

In the single atom case, \change{cf. Fig.~\ref{fig:scheme}(a), laser-cooling keeps the atomic ion at milli-Kelvin temperatures such that the atom} will remain close to the trap minimum. Consequently, we assume the atom all the time to be in the centre of the trapping potential. Furthermore we assume that the energy is not fixed, but rather given by a thermal distribution with $E$ its mean value.  Then, \change{for an} isotropic trapping potential, the distribution of impact parameters is given by
\begin{equation}\label{eq:fSAMichael}
 f_{SA}(b) = \frac{b}{\sigma^2}e^{-\frac{b^2}{2\sigma^2}}\,,
\end{equation}
where $ \sigma = \sqrt{E/\left(\mu \omega^2\right)}$ is 
the effective length of the trap at a given energy \change{and} trap frequency $\omega$.

In the second scenario, that of a large Coulomb crystal, \change{cf. Fig.~\ref{fig:scheme}(b),} the lattice spacing $d$  determines the maximum impact parameter in a scattering event, i.e., $b_{max}=d/2$. Assuming the molecular ion is entering along the [100] direction of the simple cubic lattice, cf. Fig.~\ref{fig:scheme}(b), we approximate the impact parameter distribution by 
\begin{equation}\label{eq:Michael9}
  f_{CC}(b) = \frac{8b}{d^2}, \quad b \in\left[0,\frac{d}{2}\right]\,.
\end{equation}

\begin{figure}[tb]
 \centering
\includegraphics[width=1.0\linewidth]{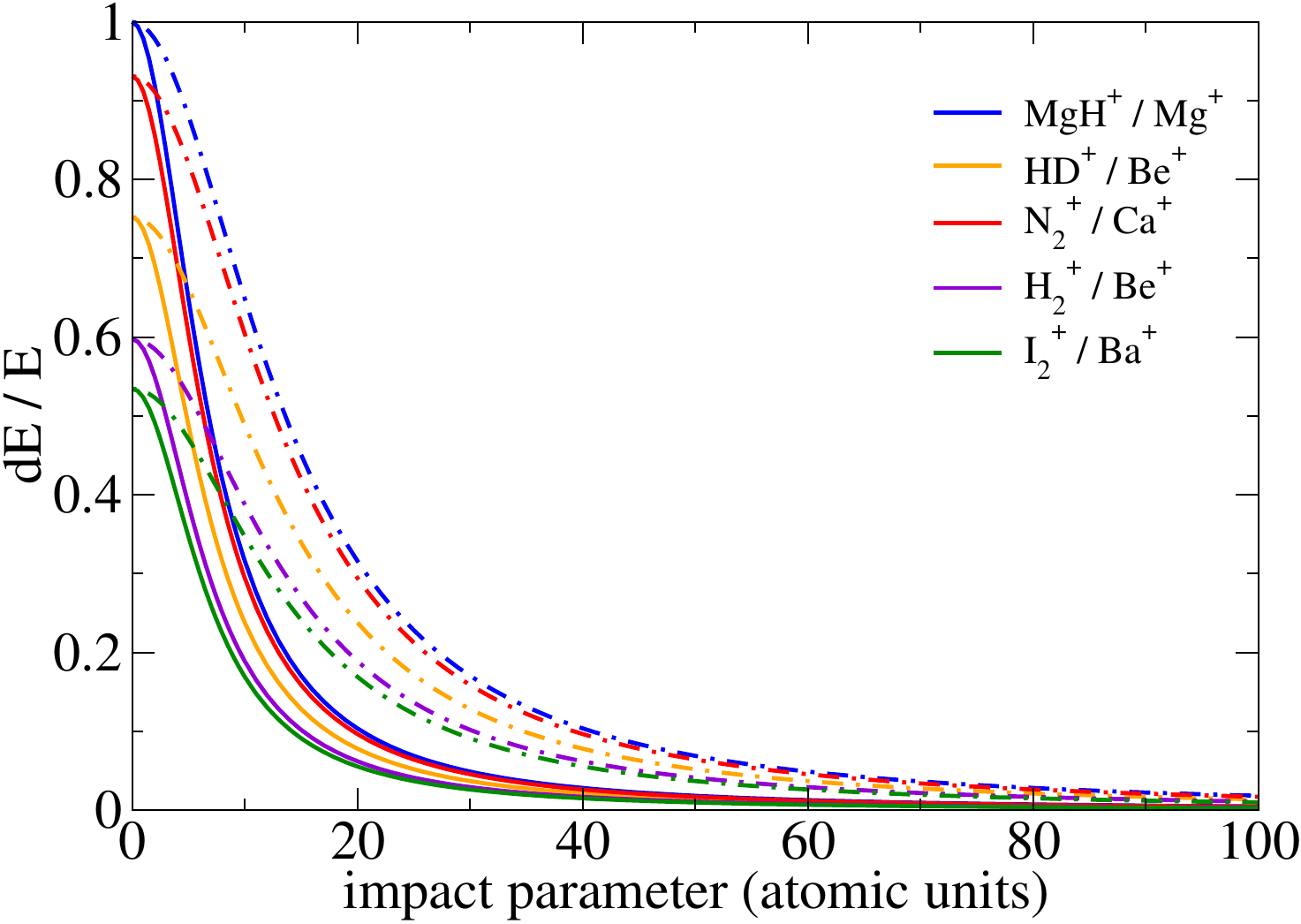}
 \caption{CM-frame translational energy transfer, relative to the collision energy, as a function of the impact parameter $b$ for two collision energies $E = 2$eV (solid lines) and $E = 1$ eV (dashed dotted lines) and various pairings of molecular and atomic ions.}
 \label{fig:dE}
\end{figure}
In the laboratory frame, the  translational energy transferred from the molecular ion to the atom in a single collision, $\delta E_{lab}$, is given by
\begin{equation}
  \label{eq:trans}
  \delta E_{lab} = \frac{2\xi(1-\cos{\theta_{sc}})}{(1+\xi)^2}E_{lab}\,,
\end{equation}
where $E_{lab}$ is the initial collision energy, $\xi = M_{mol}/M_{at}$ is the molecule to atomic mass ratio, and $\theta_{sc}$ is the scattering angle. The latter depends on the scattering energy $E$ (in the center-of-mass (CM) frame), the impact parameter $b$, and the charges $q_{at}$, $q_{mol}$ of the atomic and molecular ions respectively~\cite{BerglundPRA1}, 
\[
 \theta_{sc}(E,b) = 2\sin^{-1}{\left(1/\sqrt{1+\left(\frac{2Eb}{q_{at}q_{mol}}.\right)^2}\right)}\,.
\]
\change{For elastic collisions,} the CM and lab frame energies are \change{simply } related by 
\begin{equation}\label{eq:ELabCM}
 E_{lab} = E M_{mol}/\mu\,,
\end{equation}
\change{which} allows us to express the energy transfer \change{in the lab frame,} Eq.~\eqref{eq:trans}, to that in the CM frame. \change{The resulting CM-frame} energy transfer \change{in a single} collision, relative to the collision energy, \change{is shown in}  Fig.~\ref{fig:dE} as a function of the impact parameter $b$. \change{As one would expect,} the energy transfer \change{in a single collision} is seen to be most efficient for collision partners with similar masses. 
In addition, \change{for non-head-on collisions, the energy transfer is larger  at lower collision energies and for very large impact parameters, the relative energy transfer behaves as $\propto b^{-2}$. In this regime, the rotational excitation scales as $b^{-4}$ for polar and as $ b^{-6}$ for apolar molecules, as we show in Appendix~\ref{sec:highb}. This scaling is beneficial for keeping excitations in sympathetic cooling low. Large values of the impact parameter are relevant since the lattice spacing is typically of the order of $\mu$m.}

A complete cooling process consists of a series of such collisions, and will successively lower the translational energy of the molecular ion. Let the initial and final translational energies of the molecular ion be $E_{max}$ and $E_{min}$, respectively. For a given scattering energy, the typical transfer of kinetic energy from the molecular ion to the coolant, $\Braket{\delta E}$, is given by the average with respect to the impact parameter,
\begin{equation}\label{eq:dEMichael}
 \Braket{\delta E(E)} = \int_0^{\infty}\delta E(E,b)\,f(b)\,db\,,
\end{equation}
where the probability distribution $f(b)$ depends on the 
specific cooling scenario \change{and is found in Eqs.~\eqref{eq:fSAMichael} and~\eqref{eq:Michael9}} for the single atom and crystal cooling scenarios.
The total energy transferred in a cycle of $N$ collisions is then
\begin{equation}\label{eq:Etrans1}
 \Delta E = E_{max} - E_{min} = \sum_{n=0}^{N-1}\Braket{\delta E\change{(E_n)}},
\end{equation}
where $E_{n+1} = E_n - \Braket{\delta E(E_n)}$. If the energy transfer in each collision is very small compared to the collision energy, then the energy transfer is approximately constant over several collisions, $n(E_i)$. It is convenient to partition the full energy interval ($E \in [E_{max},E_{min}]$) into $N$ sub-intervals such that
\begin{equation}
 E_{max} = E_0 > E_1 > ... \, E_i > E_{i+1} > ... \, E_N = E_{min}\,,
\end{equation}
and, assuming that the energy interval $\Delta E_i \equiv E_i - E_{i+1}$ is not too large,
\begin{equation}\label{eq:ndef}
 \Delta E_i \approx n(E_i)\Braket{dE\change{(E_i)}} \Rightarrow n(E_i) \approx \frac{\Delta E_i}{\Braket{\delta E\change{(E_i)}}},
\end{equation}
thus defining $n(E_i)$. We can now rewrite Eq.~\eqref{eq:Etrans1} as
\begin{equation}\label{eq:Etrans2}
 \Delta E = \sum_{i=0}^{N-1}\Delta E_i \approx \sum_{i=0}^{N-1}n(E_i)\Braket{\delta E\change{(E_i)}}.
\end{equation}

We now \change{estimate the number of collisions necessary to reach a desired final energy}, beginning with the Coulomb crystal scenario, where, by Eq.~\eqref{eq:Michael9}, the mean energy loss becomes
\change{(specializing to the case of unit charges $q_aq_m = e^2$)}
\begin{eqnarray}\label{eq:Michael10}
\begin{aligned}
 \Braket{\delta E(E)}_{CC} &= \frac{8}{d^2}\int_0^{\frac{d}{2}}\delta E(E,b)bdb,  \\ 
 &=\frac{4\change{e^4}\xi\log{\left(\left(\change{\frac{d\cdot E}{e^2}}\right)^2 + 1\right)}}{(1+\xi)^2d^2E}\,.
  \end{aligned}
\end{eqnarray}
With Eqs.~\eqref{eq:ndef} and~\eqref{eq:Michael10}, we \change{can express} the number of collisions required to reduce the collision energy from $E_i$ to $E_{i+1}$ as
\begin{equation}\label{eq:ncc}
 n(E_i)_{CC} \approx \frac{(1+\xi)^2d^2E_i}{4\change{e^4}\xi\log{\left(\left(\change{\frac{d\cdot E_i}{e^2}}\right)^2 + 1\right)}}\Delta E_i\,.
\end{equation}
\change{Moreover, we} can estimate the time between collisions, 
\begin{equation}\label{eq:Michael11}
 \tau_{CC} = \frac{d}{v_{lab}} = d \sqrt{\frac{M_{mol}}{2E_{lab}}} = d \sqrt{\frac{\mu}{2E}},
\end{equation}
where $v_{lab}$ is the speed of the molecular ion in the laboratory frame \change{such that} the cooling time for a given sub-interval, $i$ is given by the product of  Eqs.~\eqref{eq:ncc} and~\eqref{eq:Michael11}. Summing over all sub-intervals leads to the total cooling time,
\begin{equation}\label{eq:Michael12}
 \begin{aligned}
 T_{CC} &\approx \frac{(1+\xi)^2d^3}{4\change{e^4}\xi}\sqrt{\frac{\mu}{2}}
 \sum_{i=0}^{N-1}\frac{\sqrt{E_i}}{\log{\left(\left(\change{\frac{d\cdot E_i}{e^2}}\right)^2 + 1\right)}}\Delta E_i \\
 &\to \frac{(1+\xi)^2d^3}{4\change{e^4}\xi}\sqrt{\frac{\mu}{2}} \int_{E_{min}}^{E_{max}}\frac{\sqrt{E}dE}{\log{\left(\left(\change{\frac{d\cdot E}{e^2}}\right)^2+1\right)}} \,,
 \end{aligned}
\end{equation}
where the integral corresponds to the limit $\Delta E_i \to dE$.

In the \change{case of cooling with a single atomic ion,  the trap width} $\sigma$ in Eq.~\eqref{eq:fSAMichael}
is large compared to any $b$ at which energy transfer is significant.  We thus find
\begin{eqnarray}
\nonumber
 \Braket{\delta E(E)}_{SA} &=& \frac{1}{\sigma^2}\int_0^{\infty}\delta E(E,b)be^{-\frac{b^2}{2\sigma^2}}db,  \\ 
  &<& \frac{\change{e^4}\xi\log{\left(\left(\change{\frac{2\sigma E}{e^2}}\right)^2 + 1\right)}}{(1+\xi)^2\sigma^2E}.
\label{eq:dEavg}
\end{eqnarray}
Using Eq.~\eqref{eq:dEavg}, the number of scattering events required to change the translational energy of the molecular ion by $\Delta E_i$ becomes
\begin{equation}\label{eq:nSI}
\begin{aligned}
 n_{SA}(E_i) &\approx \frac{\Delta E_i}{\Braket{\delta E(E_i)}}_{SA}\\
 &> \frac{(1+\xi)^2\sigma^2E_i}{\change{e^4}\xi\log{\left(\left(\change{\frac{2\sigma E_i}{e^2}}\right)^2 + 1\right)}}\Delta E_i.
 \end{aligned}
\end{equation}
Since the molecular ion oscillates in the trap \change{back and forth whereas the atom remains at the trap center}, the time between two collisions amounts to $\tau_{SA} = \pi/\omega$, independent of $E$. By this approximation we essentially neglect the fact that the Coulomb scattering cross section is infinite. Discretizing the  energy range, the total time needed to lower the molecular ion's energy by $\Delta E$ can be approximated by
\begin{equation}\label{eq:Michael7}
\begin{aligned}
  T_{SA} &\approx \sum_{i=0}^{N-1}
             n_{SA}(E_i)\tau_{SA} 
  = \frac{\pi}{\omega}\sum_{i=0}^{N-1} 
n_{SA}(E_i) \\
        &> \pi\sqrt{\mu}\sum_{i=0}^{N-1}
  \frac{(1+\xi)^2\sigma_i^3\sqrt{E_i}}{\change{e^4}\xi\log{\left(\left(\change{\frac{2\sigma_iE_i}{e^2}}\right)^2 + 1\right)}}\Delta E_i
  \end{aligned}
\end{equation}
with $\sigma_i = \sqrt{E_i/\mu\omega^2}$. From Eq.~\eqref{eq:Michael7}, it is clear that cooling at high energies is much slower than at low energies. As a matter of fact, one can estimate an average value for $\tilde{\sigma} \approx 6\sigma(E_{max})/7$, cf. Appendix~\ref{sec:sigmaavg}, which allows us to approximate the total cooling time, 
\begin{equation}\label{eq:Michael8}
 \begin{aligned}
  T_{SA} > \pi\sqrt{\mu}\tilde{\sigma}^3
  \sum_{i=0}^{N-1}\frac{(1+\xi)^2\sqrt{E_i}}{\change{e^4}\xi
  \log{\left(\left(\change{\frac{2\tilde{\sigma} E_i}{e^2}}\right)^2 + 1\right)}}\Delta E_i\,.
 \end{aligned}
\end{equation}

\change{Comparing Eqs.~\eqref{eq:Michael8} and~\eqref{eq:Michael12}, we find}
a simple relation between the cooling times in the two regimes,
\begin{equation}\label{eq:Tratio}
 \frac{T_{SA}}{T_{CC}} \sim \left(\frac{\tilde{\sigma}}{d}\right)^3\,.
\end{equation}
For typical Coulomb crystals, $d \approx 10\,\mu$m,  whereas $\sigma(E_0) = 635\,\mu$m for $E = 2\,$eV. With 
$\omega = 2\pi\times 1\,$MHz, $T_{SA}$ is more than $10^{6}$ times larger than $T_{CC}$.
As an example, cooling $^{24}$MgH$^+$ from $2\,$eV to $0.01\,$eV in a crystal
of $^{24}$Mg$^+$ with $d=5.29\,\mu$m 
takes approximately  $2\,$ms in agreement with an earlier estimate~\cite{BussmannIJMS06}. This needs to be 
compared to $\sim$$40\,$min when cooling with a single atomic ion. Sympathetic translational cooling is thus much more advantageous with a Coulomb crystal.

In order to validate our model for sympathetic cooling in a Coulomb crystal, we compare our estimated cooling times with cooling times obtained from molecular dynamics (MD) simulations. The comparison is presented in Table~\ref{tab:comp}.
\change{While we are mainly interested in singly charged ions, we can evaluate our estimates also for the higher charges used in the MD simulations for the sake of the comparison.  
Remarkably, our estimates are of the same order of magnitude, even for charges as high as $40$. For smaller charges the agreement of our estimates with the MD simulations is even better. Such better agreement with decreasing charge is expected since collective effects (that are accounted for in the MD simulations but not in our model) should be less pronounced. Given the many simplifications and approximations of our model, the good agreement with full MD simulations is encouraging. It  suggests that our model provides, at least for small charges, a simple means of assessing whether a sympathetic cooling experiment  is worth pursuing, assuming that one wishes to avoid planning for extra efforts on rotational cooling.}
\begin{table}[tb]
  \centering
  \begin{tabular}{|c| r| r| r| r|}
  \hline
   $n$ & $\braket{\delta E_{lab}}$ (eV) & $n_{coll}$ & 
    $T_{CC}$ (ms) & $T_{MD}$* (ms)\\
    \hline
    \hline
    $10$ & $1.15\cdot10^{-5}$ & $3.5\cdot10^{4}$ & $3.5\cdot10^{-1}$ & $\approx4.6\cdot10^{-1}$ \\
    $20$ & $4.80\cdot10^{-5}$ & $8.3\cdot10^{3}$ & $8.5\cdot10^{-2}$ & $\approx1.3\cdot10^{-1}$ \\
    $30$ & $1.14\cdot10^{-4}$ & $3.5\cdot10^{3}$ & $3.6\cdot10^{-2}$ & $\approx6.0\cdot10^{-2}$ \\
    $40$ & $2.11\cdot10^{-4}$ & $1.9\cdot10^{3}$ & $1.9\cdot10^{-2}$ & $\approx3.6\cdot10^{-2}$ \\
    \hline
  \end{tabular}
  \caption{Comparison of the cooling time in the Coulomb crystal scenario, as obtained by molecular dynamics (MD) simulations and by our model. The comparison uses the scattering pair $^{24}$MgH$^+$-$^{24}$Mg$^{n+}$, where $n$ is the charge of the atomic ion, a starting energy $E_{max} = 0.4$ eV and $E_{min} = 0.0$ eV, in the lab frame. In the flat distribution, Eq.~\eqref{eq:Michael9}, the cutoff is $d = 17.5$ $\mu$m.
  The collision rate was estimated from the initial speed and mean free path and found to be $\approx 280$ MHz.
  $T_{CC}$ was calculated using one term in the series expansion in Eq.~\eqref{eq:Michael12}, using the listed values of the average energy transfer in a single collision, $\Braket{\delta E_{lab}}$,  and of the number of collisions, $n_{coll}$, obtained from Eq.~\eqref{eq:Michael10} and Eq.~\eqref{eq:ncc}, respectively. *$T_{MD}$ corresponds to the time reported  in Ref.~\cite{BussmannIJMS06}.}
  \label{tab:comp}
\end{table}

One may wonder if local disruption of the crystal structure and heating effects due to collisions may have a significant effect on subsequent collisions. However, since the speed of sound is typically of the order of $\sim10\,$m/s, the ions will during most of the cooling process be moving with supersonic velocities. Therefore, previous collisions have no effect as long the crystal is long enough to stop the incoming ion by passing through once. Even if this is not the case, the mean density of cold ions will not change dramatically after passing of a fast ion, and the crystal will be essentially the same before the next passage. 

\section{Rotational excitation during the cooling process}
\label{sec:rot-excitation}
Just as with the kinetic energy transfer, the rotational excitation depends on the scattering energy and the impact parameter. Let the rotational excitation \change{associated with a single collision} at a given energy and impact parameter, \change{obtained in Ref.~\cite{BerglundPRA1}}, be $\epsilon(E,b)$.   In order to estimate the rotational excitation over a complete cooling cycle, we discretize the range from the initial to the final scattering energy, analogously to Eq.~\eqref{eq:Michael12}. 
Just like in the treatment of the translational cooling, we need to average over the impact parameter. We are now only concerned with the crystal cooling scenario, so we use the distribution function in Eq.~\eqref{eq:Michael9}, obtaining $\tilde{\epsilon}(E) = \Braket{\epsilon(E,b)}_{CC}$. If the accumulated cycle excitation is to be small, then certainly $\epsilon(E,b)$ need be small, and since most excitation takes place for impact parameters much smaller than $d$~\cite{BerglundPRA1}, we must have $\tilde{\epsilon}(E) \ll 1$. Once we have obtained $\tilde{\epsilon}(E_i)$ corresponding to the energy of subinterval $i$, we can estimate the accumulated rotational excitation on this interval by $n(E_i)\tilde{\epsilon}(E_i)$. The accumulated rotational excitation in a complete cycle is obtained by summing up the contributions from the subintervals
\begin{equation}\label{eq:Sigma}
 \Phi_{\Sigma} \approx \sum_{i=0}^{N-1}n(E_i)\tilde{\epsilon}(E_i)\,.
\end{equation}
Notice that Eq.~\eqref{eq:Sigma} disregards de-excitations that might occur in a multiple scattering sequence. However, this effect will be minor as long as the degree of excitation during the complete translational cooling cycle is much smaller than 1 as assumed above.  %

In order to obtain the accumulated excitation we need to first to estimate the single collision excitation, $\tilde{\epsilon}(E)$. Since it fundamentally depends on whether it is a polar and apolar molecular ion that is sympathetically cooled, in the following we consider the two cases separately.  
\subsection{Apolar molecular ions}
\begin{figure}[tb]
 \centering
\includegraphics[width=\linewidth]{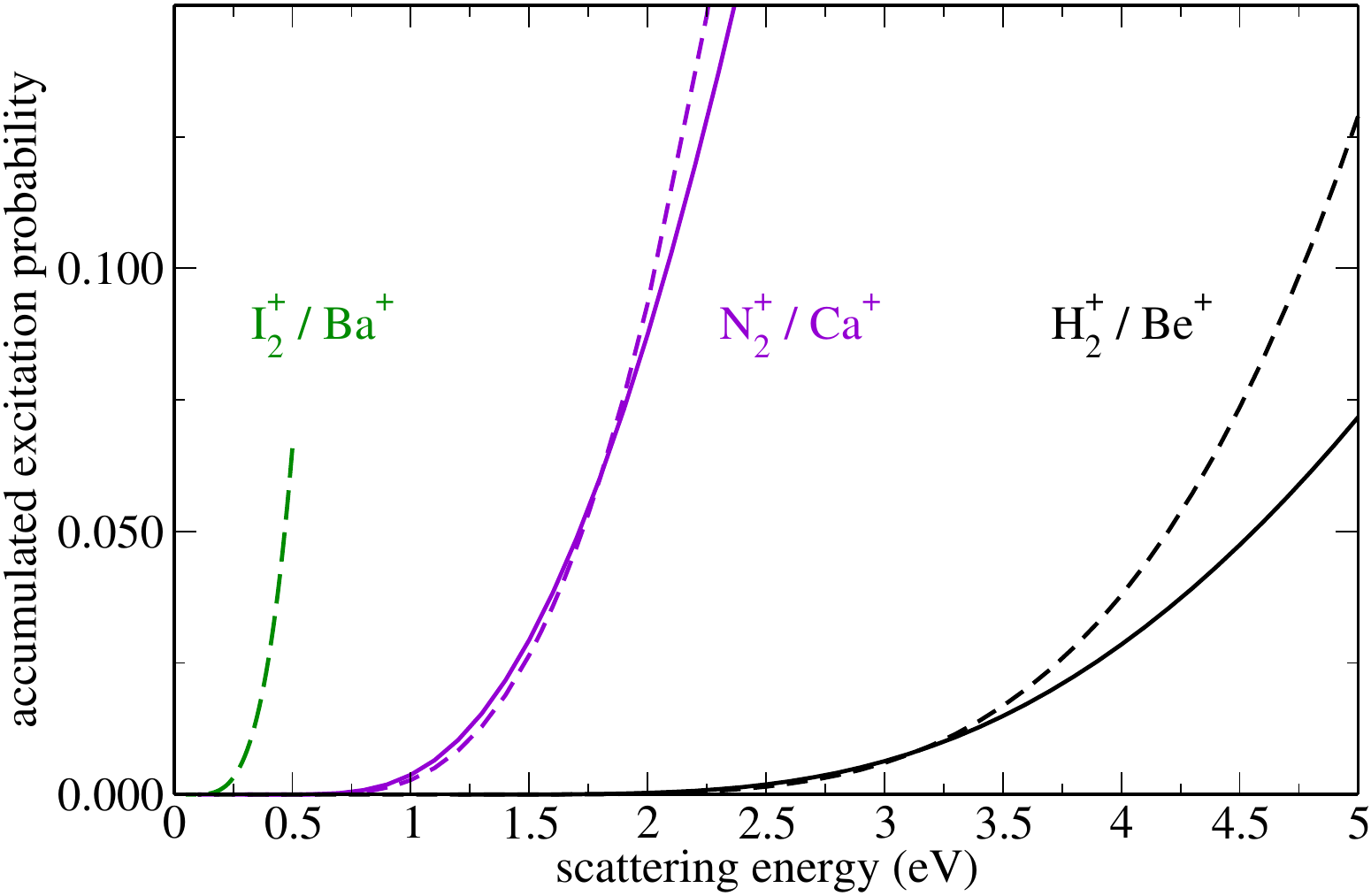}
 \caption{Accumulated excitation for apolar molecular ions after a complete cooling cycle as function of the initial scattering energy, comparing \change{numerical calculations taking both the quadrupole and polarizability interactions into account~\cite{BerglundPRA1}} (dashed lines) to 
PT (solid lines, $d \approx 5.3$ $\mu$m).}
 \label{fig:etapop}
\end{figure}
For apolar molecular ions, we \change{have} calculated numerically the population excitation \change{in a single collision,} taking the polarizability and quadrupole interactions into account, \change{and only the stronger quadrupole interaction for the PT-treatment}~\cite{BerglundPRA1}. 
For a full cooling process we need to calculate the accumulated rotational excitations by averaging over the impact parameter. In particular, we are interested in calculating averages for the excitations in a single collison at a fixed scattering energy. In our model, the only quantity in the expression for the single collision excitation that depends on the impact parameter is the third power of the maximum field strength~\cite{BerglundPRA1}, 
\begin{equation}\label{eq:epsb}
\varepsilon_0^3(E, b) = \frac{1}{\left(\frac{\change{e^2}}{2E} + \sqrt{\left(\frac{\change{e^2}}{2E}\right)^2 + b^2}\right)^6}.
\end{equation} 
In the CC-scenario, we use Eqs.~\eqref{eq:Michael9} and~\eqref{eq:epsb} to calculate the average
\begin{equation}\label{eq:avgeps}
 \frac{8}{d^2}\int_0^{\frac{d}{2}}\varepsilon_0^3(E, b)b\,db 
 \approx \frac{6}{5d^2}\change{\left(\frac{E}{e^2}\right)^4} \,.
\end{equation}
where the evaluation is presented in Appendix~\ref{sec:eps0avg}. From this average we have now obtained an estimate for the population excitation at a given scattering energy in terms of the reduced mass, the lattice spacing, the rotational constant, $B$ and the quadrupole moment along the molecular axis, $Q_Z$,
\begin{widetext}
 \begin{eqnarray}\label{eq:exc_quad}
  \tilde{\epsilon}(E) \approx \left\langle \left|c_{2,0}^{(1)}(E)\right|^2\right\rangle &\approx& 1.86^2\frac{1}{150d^2}\left(\frac{3Q_Z}{4}\right)^2\change{e^4}\mu E
  \left(1+ 6 \cdot 1.86 \sqrt{\frac{\change{e^4}\mu}{E^3}}B\right)e^{-6 \cdot 1.86\sqrt{\frac{\change{e^4}\mu}{E^3}}B}.
 \end{eqnarray}
\end{widetext}
This expression for the typical rotational excitation in a single collision, together with Eqs.~\eqref{eq:ncc} and~\eqref{eq:Sigma} are what we need to estimate the accumulated cycle excitation. If we let the energy partition become infinitesimal, i.e. let $\Delta E_i \to dE$ in Eq.~\eqref{eq:ncc}, then the sum in Eq.~\eqref{eq:Sigma} becomes an integral, 
\begin{eqnarray}
  \label{eq:cycleq_ana}
  \Phi_{\Sigma} &\approx&  1.86^2\frac{3(1+\xi)^2\mu Q_Z^2}{200\xi}\times
  \\&&\int_{E_i}^{E_f}\frac{E^2\left(1 + 6 \cdot 1.86\sqrt{\frac{\change{e^4}\mu}{E^3}}B\right)
  e^{-6 \cdot 1.86\sqrt{\frac{\change{e^4}\mu}{E^3}}B}}{\log{\left(\left(\change{\frac{d\cdot E}{e^2}}\right)^2+1\right)}}dE\,.
       \nonumber 
\end{eqnarray}
which is our final result for apolar molecular ions.
The final estimate for the accumulated excitation probability at the end of the cooling process nearly only depends on the molecular parameters and initial scattering energy, where the dependence on the lattice splitting is weak due to the logarithm.

In Fig.~\ref{fig:etapop}, numerical and analytical results are presented for a few selected combinations of molecular and atomic ion species. For the popular example of N$_2^+$ - Ca$^+$~\cite{TongPRL10,GermannNatPhys14,GermannJCP16a,GermannJCP16b}, we expect excitation of more than a few percent only for initial scattering energies well above 1.5$\,$eV (CM). However, for very heavy molecular ions with small rotational constants, such as I$_2^+$, significant rotational excitation is expected already for initial energies of a few hundred meV. At the other end we find the scattering pair H$_2^+$ - Be$^+$, representing a molecular ion with a large rotational constant. In this case we observe a resilience to rotational excitations, our results indicate that significant excitations above a few procent to occur above 3.5$\,$eV (CM). \change{Notice} that the discrepancy between analytical and numerical results for higher energies are more significant for H$_2^+$ than for N$_2^+$. The reason for this is our neglect of the polarizability interaction for the analytical calculations. The polarizability becomes more important for high energies, and the polarizability interaction is relatively more \change{pronounced} for H$_2^+$~\cite{BerglundPRA1}. 
In Fig.~\ref{fig:etapop}
we have assumed $d = 5.3$ $\mu$m~. However, since the main contribution to both the sympathetic cooling process and the rotational excitations take place at very much shorter distances, neither d nor actually the cooling scenario are important for the complete cooling cycle result.   
The total cooling time, however,  strongly depends on the particular scenario and values of $d$ as explained above.
\subsection{Polar molecular ions}
\begin{figure}[t!]
 \centering
\includegraphics[width=\linewidth]{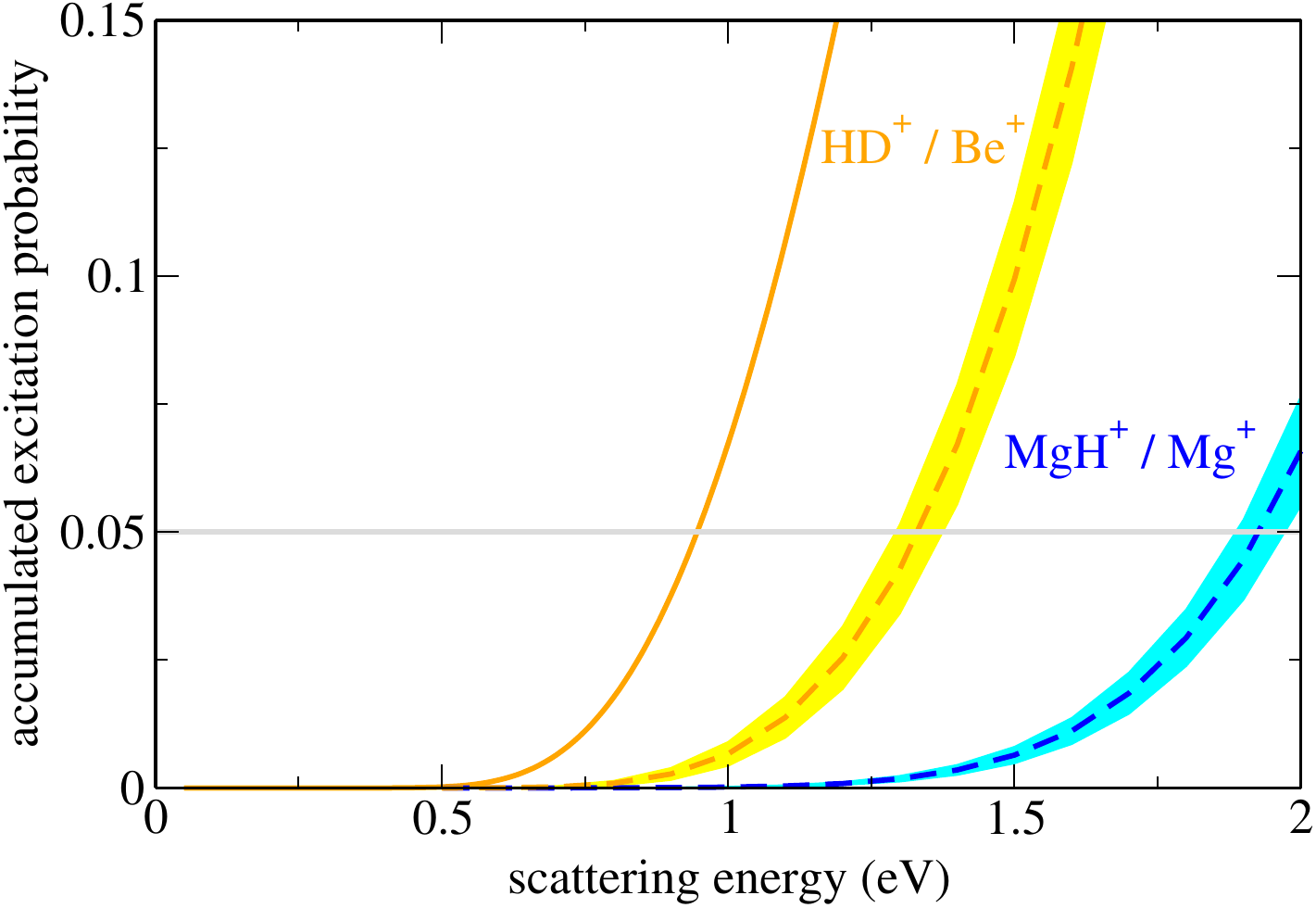}
 \caption{Accumulated excitation probability, Eq.~\eqref{eq:Sigma}, for polar molecular ions after a complete cooling cycle, as a function of 
 the initial scattering energy with $E_{final} = 0.1\,$eV, comparing the estimate (solid lines) based on approximation~\eqref{eq:cAdiabatic} 
 (with $\delta E = 0.05\,$eV) to full numerical simulations (dashed lines, $\delta E = 0.1$~eV).
Within each interval $\delta E$, the excitation occurring in a single collision, 
$\tilde{\epsilon}$, can be evaluated at the highest and lowest energy, $E_i$ and $E_i -\delta E$, defining the shaded region, or taken to be the arithmetic mean, indicated by the dashed lines.
The horizontal gray line marks an excitation level of 5\%.}
 \label{fig:popchi}
\end{figure}
For the polar case, the dominant interaction between the molecular ion and the electric field originating on the coolant atomic ion is the dipole interaction, which is linear in the electric field strength. This combined with the gradual variation of the Coulomb interaction at large distances leads to a slow onset (offset) of the interaction between the two scattering ions as they approach (receed). This in turn allows for the single collision dynamics to be analyzed in the adiabatic picture, as done in Ref.~\cite{BerglundPRA1}. In general, the scattering is in the high field limit, leading to strong following of the field, and no closed formula can be found for $\tilde{\epsilon}(E)$. However, for molecular ions having a low permanent dipole moment, the effect of the field on the rotational states is moderate, and an approximate, two-level model can be used, which leads to the following expression for $\epsilon(E_i)$~\cite{BerglundPRA1}:
\begin{widetext}
\begin{equation}\label{eq:cSingle}
 \tilde{\epsilon}(E) = \frac{(1.86\pi D)^2\change{e^4}\mu}{6E^3b_{max}^2}\int_0^{b_{max}}
 \frac{\exp{\left(-2\cdot 1.86B\sqrt{\frac{\change{e^4}\mu}{E^3}}\sqrt{1 + \frac{D^2}{3B^2\left(\frac{\change{e^2}}{2E} + \sqrt{\left(\frac{\change{e^2}}{2E}\right)^2 + b^2}\right)^4}}\right)}}{\left(\frac{\change{e^2}}{2E} + \sqrt{\left(\frac{\change{e^2}}{2E}\right)^2 + b^2}\right)^4 + \frac{D^2}{3B^2}}
 b\,d b.
\end{equation}
\end{widetext}

From this we estimate the accumulated excitation in a complete cooling cycle using Eq.~\eqref{eq:Sigma} to be
 \begin{equation}\label{eq:cAdiabatic}
 \begin{aligned}
  \Phi_{\Sigma} &\approx \sum_{i=1}^N n(E_i)\tilde{\epsilon}(E_i) \\
  &\approx \frac{\left(1 + \xi \right)^2d^2}{4\change{e^4}\xi}\sum_{i=1}^N\frac{E_i\tilde{\epsilon}(E_i)}{\log{\left(\change{\frac{d\cdot E_i}{e^2}}\right)^2 + 1}}\Delta E_i
  \end{aligned}
 \end{equation}

The accumulated excitation is presented in Figure~\ref{fig:popchi}. Here we present both analytical and numerical results for HD$~+$, representing the low-field limit, and only numerical results for MgH$^+$,representing the high-field limit. For the low-field limit (orange) our estimate, Eq.~\eqref{eq:cAdiabatic} is in fair agreement with the numerical results, but not as well as for the apolar case. We notice that our estimate gives an upper bound to the numerical results at relevant scattering energies. We attribute this effect, in part, because single collisions tends to give an upper bound for low energies and high values of the impact parameters, where the effect of the field is weak~\cite{BerglundPRA1}. Notice also that the averaging process, Eq.~\eqref{eq:cSingle} suppresses the value of the integral at low impact parameters, due to the differential elemment, $b\, d b$. This resoning fails at higher energies, as can be seen in Figure~\ref{fig:popchi}, but at such high energies the probability excitation is so high that an estimate is no longer relevant. In the high-field limit we have no estimate for the single excitation~\cite{BerglundPRA1}, and we can therefore not estimate the full cycle either. A high dipole moment does suppress the rotational excitation process~\cite{BerglundPRA1}, a fact that is reflected in the low accumulated excitation probability seen in the figure, and we can therefore expect, contraintuitively, that a molecular ion with a high permanent dipole moment should be resilient against rotational excitations in a sympathetic cooling process.
The accumulated excitations obtained from numerical calculations of single collision excitations predict a total excitation of $0.05$ at $E \approx 1.8, 1.25$~eV for MgH$^+$ and HD$^+$ respectively. By Eq.\eqref{eq:ELabCM} these energies correspond to $E_{lab} \approx 3.5, 1.6$~eV for MgH$^+$ and HD$^+$ respectively. For the apolar species N$_2^+$, H$_2^+$ and I$_2^+$ the corresponding lab energies are $E_{lab} \approx 2.6, 4.9, 2.8$~eV.

One may wonder if the trapping field also could lead to rotational excitations. The strongest fields are found in rf traps, where the dominating rf-fields can reach  maximum amplitudes of $\sim10^5$V/m. For typical values dipole moments of polar diatomic molecular ions, such fields leads to a coupling energy of $\sim10^{-24}$ Joule or equivalent a frequency of $\sim 1$ GHz. Since this frequency is orders of magnitude lower than the frequencies associated with the rotational splitting in the molecular ions considered above, we do not expect any significant contribution of the trapping fields to rotational state excitations. Notice however that even though the trap field is negligible in terms of the rotational state splitting it can lead to mixing of $m_J$-sub-levels~\cite{HashemlooJCP15, HashemlooIJMPC16}.

\section{Conclusions and outlook}
\label{sec:concl}
\change{We have constructed an approximation that allows us to estimate the overall rotational excitation accumulated over many collisions in the sympathetic cooling of a diatomic molecular ion by laser-cooled atomic ions.
We have considered two possible experimental scenarios, that of cooling with a single atomic ion and that of a Coulomb crystal of trapped atomic ions, as well as polar and apolar molecules.
With regards to the rotational excitation at the end of the sympathetic cooling,} we have found it to be considerably different for apolar and polar molecular ions. \change{For polar molecules, an adiabatic picture describes the rotational dynamics best~\cite{BerglundPRA1} which yields an} estimate that is conservative and not very accurate, in particular for high collision energies corresponding to high field scattering. For polar molecular ions an accurate estimate requires a full quantum-dynamical treatment, whereas validity of perturbation theory for apolar molecular ions has allowed us to derive a closed-form estimate of the accumulated population excitation which solely depends on the molecular parameters and initial scattering energy~\cite{BerglundPRA1}. \change{This then translates also into the estimates of the overall rotational excitation.}
The scenarios of using a Coulomb crystal of atomic ions or just a single atomic ion do not significantly change the final degree of collision-induced rotational excitation. However, translational cooling with a single atomic ion is dramatically slower and will generally be impractical.

\change{More precisely,}
for a wide range of apolar molecular ions, we find the internal state to be preserved for initial energies of 1$\,$eV and above, eventually limited by close-encounter interactions disregarded in the present treatment. Our results should be of interest for \change{designing of new cooling experiments. In particular, for a given molecular ion of interest, our estimates can serve as a guide for choosing the trap depth that allows for  keeping the rotational excitations below a desired} threshold. Conversely, if the potential depth is set for a given experiment, our results can serve as a guide indicating which molecular ions are \change{suitable for sympathetic cooling without suffering rotational state changes}.

\change{There is presently growing interest in polyatomic molecules~\cite{HutzlerQST20,ErezPRX23,CalvinPRL25}}.
When extending \change{our} treatment to polyatomics, 
we expect rotational excitation to be more critical, both because of  more degrees of freedom with low-energy spacings and the physical size of the molecules making close-encounter interactions more likely. 
The latter deserve a more thorough investigation in future work as they might provide a new avenue for controlling collisions due to the extremely large fields present in a close encounter. 
The control knob would be the initial collision energy which can be varied  via the choice of the molecule's position in the trap during photo-ionization, or by injecting low-energetic molecular 
ions from an external source into the trap. 
The same techniques could also be used to experimentally test our present predictions for diatomics.

\begin{acknowledgments}
We would like to thank Stefan Willitsch for fruitful discussions on the quadrupole aspect of the work. 
Financial support from the State Hessen Initiative for the Development of Scientific and Economic Excellence (LOEWE), the European Commission’s FET Open TEQ, the Villum Foundation, and the Independent Research Fund Denmark is gratefully acknowledged.
This research was supported in part by the National Science Foundation under Grant No. NSF PHY-1748958.
JMB wishes to thank SECIHTI for providing a postdoc scholarship.
\end{acknowledgments}

\appendix
\section{Collisions with large impact parameter $b$}
\label{sec:highb}

The energy transfer at a scattering energy $E$ and impact parameter $b$ is given by
\begin{equation}\label{eq:et}
 \delta E(E,b) = \frac{2\rho}{\left(1+\rho\right)^2}
 \left(1 - \cos{\theta_{sc}(E,b)}\right)E,
\end{equation}
where
\begin{equation}
 \theta_{sc}(E, b) = 2\sin^{-1}{1 / \sqrt{1 + \left(\frac{2Eb}{e^2}\right)^2}}
\end{equation}
and $e$ is the elementary charge (we assume unity charges here) and $\rho = \frac{M_{mol}}{M_{at}}$. The maximum field strength is given by
\begin{equation}\label{eq:ve}
 \varepsilon_0 = \frac{1}{r_0^2} = \frac{1}{\left(\frac{e^2}{2E} + \sqrt{\left(\frac{e^2}{2E}\right)^2 + b^2}\right)^2}
\end{equation}
We are interested in estimating the behaviour of these quantities in the high $b$ limit. 
We see for apolar molecular ions the population excitation probability is $\propto \varepsilon_0^3 = \frac{1}{r_0^6}$ via perturbation theory. For polar species we can use the low field 2-level expression to obtain population excitation probability $\propto \varepsilon_0^2 = \frac{1}{r_0^4}$.

We proceed to estimate the energy transfer for high $b$. First define $x = \frac{2Eb}{e^2}$. Then for $x \gg 1$ (high $b$)
\begin{equation}
 \theta_{sc}(E, b) \approx 2\sin^{-1}{x^{-1}} \equiv \frac{1}{x} \approx \sin{\frac{\theta_{sc}}{2}} \approx \frac{\theta_{sc}}{2},
\end{equation}
since $x^{-1} \ll 1$. In other words $\theta_{sc} \approx \frac{2}{x}$. We substitute this value for $\theta_{sc}$ in Eq.~\eqref{eq:et} we get 
\begin{equation}
\begin{aligned}
 \delta E(E,b) &\approx \frac{2\rho}{\left(1+\rho\right)^2}
 \left(1 - \left(1 - \frac{\theta_{sc}(E,b)^2}{2}\right)\right)E \\
 &= \frac{2\rho}{\left(1+\rho\right)^2}\frac{4}{2x^2}E \\
 &= \frac{4\rho}{\left(1+\rho\right)^2}\frac{e^2}{4E^2b^2}E \\
 &= \frac{\rho}{\left(1+\rho\right)^2}\frac{e^2}{Eb^2} \propto b^{-2}.
 \end{aligned}
\end{equation}

For high $b$ we see from Eq.~\eqref{eq:ve} that
\begin{equation}
 \varepsilon_0 \to \frac{1}{b^2}
\end{equation}
for large $b$. This implies that the excitation probability for polar species is proportional to $b^{-4}$ and for apolar species it is proportional to $b^{-6}$. This follows from the absolute squares of Eqs. (16), (24) and (29) in the companion paper~\cite{BerglundPRA1}.

\section{An estimate of an average of $\sigma(E)$ in the single ion cooling scenario}\label{sec:sigmaavg}

The average of $\sigma(E) = \sqrt{\frac{E}{\mu\omega^2}}$ in the single ion cooling scenario, from an initial scattering energy $E_{max}$ to a final scattering energy $E_{min}$, can be estimated by
\begin{equation}\label{eq:ASigma}
 \tilde{\sigma} = \frac{1}{N}\int_{E_{min}}^{E_{max}}\sigma(E)n(E)\,dE,
\end{equation}
where the total number of collisions is 
\begin{equation}\label{eq:AN}
 N = \int_{E_{min}}^{E_{max}} n(E)\,dE
\end{equation}
and $n(E)$ is given by Eq.~\eqref{eq:nSI}, i.e. we will approximate $n(E)\,dE \approx \frac{(1+\xi)^2\sigma^2E}{\xi\log{((2\sigma E)^2 + 1)}}\,dE$. Since the logarithm changes slowly compared to other factors in the integrand we will treat it as a constant, $\frac{1}{\log{(\tilde{x}^2+1)}}$, for some $2\sigma(E_{min})E_{min} < \tilde{x} < 2\sigma(E_{max})E_{max}$. With this approximation we arrive at
\begin{equation}
 N \approx \frac{\alpha}{3}\left(E_{max}^3 - E_{min}^3\right),
\end{equation}
where $\alpha = \frac{(1+\xi)^2}{\xi}\frac{1}{\mu\omega^2}\frac{1}{\log(\tilde{x}^2+1)}$. Now, Eq.~\eqref{eq:ASigma} becomes
\begin{equation}
 \tilde{\sigma} \approx \frac{\frac{2\alpha}{7}\sqrt{\frac{1}{\mu\omega^2}}\left(E_{max}^{7/2} - E_{min}^{7/2}\right)}{\frac{\alpha}{3}\left(E_{max}^3 - E_{min}^3\right)}.
\end{equation}
Assuming that the lower energy is much smaller than the maximum, or $E_{min} \to 0$, we obtain
\begin{equation}
 \tilde{\sigma} \approx \frac{6}{7}\sqrt{\frac{1}{\mu\omega^2}}E_{max}^{1/2} = \frac{6}{7}\sigma(E_{max}).
\end{equation}

\section{Evaluation of $\Braket{\varepsilon_0^3}_{CC}$}
\label{sec:eps0avg}

The average of $\varepsilon_0^3$~\cite{BerglundPRA1} over the impact parameter in the crystal cooling scenario is
\begin{equation}
 \Braket{\varepsilon_0^3}_{CC} = \frac{8}{d^2}\int_{0}^{\frac{d}{2}}\frac{b\;db}{\left(\frac{1}{2E}+\sqrt{\left(\frac{1}{2E}\right)^2 + b^2}\right)^6}
\end{equation}
Let $r = \frac{1}{2E}$. The integral can be evaluated by a symbolic mathematical software program, such as wxmaxima, with the result
\begin{widetext}
\begin{equation}
\begin{aligned}
 \Braket{\varepsilon_0^3}_{CC} &= \frac{8}{d^2}\int_0^{\frac{d}{2}}\frac{bdb}{\left(r + \sqrt{r^2 + b^2}\right)^6} \\
 &=\frac{8}{d^2}
 \left(\frac{-5\left(\frac{d}{2}\right)^6 + \sqrt{\left(\frac{d}{2}\right)^2 + r^2}
 \left(24\left(\frac{d}{2}\right)^4r + 88\left(\frac{d}{2}\right)^2r^3 + 64r^5\right) - 60\left(\frac{d}{2}\right)^4r^2 - 120\left(\frac{d}{2}\right)^2r^4 - 64r^6}{20\left(\frac{d}{2}\right)^{10}} + \frac{3}{320r^4}\right)
 \end{aligned}
\end{equation}
\end{widetext}
With $\frac{d}{2} \gg r = \frac{1}{2E}$, the main contribution from the first term in the parenthesis is $\left| -\frac{1}{4\left(\frac{d}{2}\right)^4} \right| \ll \frac{3}{320r^4}$. Therefore,
\begin{equation}
 \Braket{\varepsilon_0^3}_{CC} \approx \frac{8}{d^2}\frac{3}{320r^4} = \frac{12}{10d^2}E^4.
\end{equation}

\bibliography{refs}
\end{document}